\documentclass[pamm,a4paper,fleqn]{w-art}
\usepackage{times,cite,w-thm}
\usepackage[T1]{fontenc}
\usepackage[utf8]{inputenc}
\usepackage{booktabs}
\usepackage{siunitx}

\usepackage{amsmath}
\usepackage[ruled,vlined,linesnumbered]{algorithm2e}
\usepackage{graphicx}
\usepackage[caption=false,font=footnotesize]{subfig}

\def\Ifp{I_\mathrm{fp}}
\def\Imem{I_\mathrm{mem}}
\DeclareSIUnit\flop{FLOP}
\begin{document}

\def\leftmark{S. Long et al.: Task-Based Parallelism for Red-Black Gauss-Seidel on 2D Grids}
\def\rightmark{Preprint}


\TitleLanguage[EN]
\title[The short title]{Exploiting Task-Based Parallelism for the Red-Black Gauss-Seidel Method on 2D Grids}

\author{\firstname{Shiting} \lastname{Long}\inst{1,}%
\footnote{Corresponding author: e-mail \ElectronicMail{shitingl@kth.se}}} 
\address[\inst{1}]{\CountryCode[SE]KTH Royal Institute of Technology}
\author{\firstname{Gustavo} \lastname{Ramirez-Hidalgo}\inst{2,}}%
\address[\inst{2}]{\CountryCode[DE]Forschungszentrum J\"ulich GmbH}
\author{\firstname{Andreas} \lastname{Frommer}\inst{3,}}%
\address[\inst{3}]{\CountryCode[DE]Bergische Universität Wuppertal}
\author{\firstname{Dirk} \lastname{Pleiter}\inst{1,4,}}%
\address[\inst{4}]{\CountryCode[NL]University of Groningen}
\AbstractLanguage[EN]
\begin{abstract}
Gauss–Seidel is a well-established iterative method for the solution of linear systems, and multicoloring has been widely used to increase parallelism in iterative solution techniques. Implementing multi-color Gauss–Seidel with conventional divide-and-conquer parallelization strategies, however, may be inefficient due to global synchronization requirements and load imbalances. Task-based programming models can mitigate these issues by enabling fine-grained parallelism, removing global barriers and allowing updates of different colors to partially overlap in time. 

In this work, we implement the red–black Gauss–Seidel method using two task-based programming models and compare them with a classical divide-and-conquer parallel implementation to evaluate the impact of fine-grained parallelism on execution efficiency. The red–black scheme serves as a representative example, as task-based approaches naturally extend to more general multi-color schemes arising from unstructured grids and wider stencils. Using the solve of the 2D Poisson equation as benchmark, our results show that task-based implementations can achieve performance comparable to conventional divide-and-conquer parallelization while providing greater resilience to hardware-level asynchronicity.
\end{abstract}
\maketitle                   
\thispagestyle{empty}

\section{Introduction}

Contemporary performance improvements are increasingly driven by parallel computing rather than higher CPU clock frequencies. As a result, achieving better performance depends on decomposing workloads into concurrent tasks, enabling efficient utilization of hardware resources and translating computational capacity into measurable speedups.

Programmers are therefore required to exploit the performance potential of modern many-core processors through node-level and/or thread-level parallelism. Since parallelization cannot be handled entirely by compilers, applications must be adapted to these architectures by explicitly defining task granularity for distributing computation across cores, as well as synchronization mechanisms to ensure data consistency among parallel tasks.

The \textit{de facto} parallelization approach for numerical applications is to decompose or transform the problem into smaller partitions, where solving each partition constitutes a task that can be executed in parallel with the others. This divide-and-conquer strategy maps naturally onto the fork-join multithreading programming model, enabling efficient utilization of many-core systems and achieving parallel speedup.

The divide-and-conquer strategy is not suitable for all algorithms. One example is block LU factorization, where the number of concurrently executable block updates varies throughout execution. When using a fork-join model, threads may spend significant time waiting at synchronization barriers, resulting in hardware underutilization. In such cases, a programming model that supports asynchronous execution and dynamic work scheduling becomes essential.
Furthermore, modern computing systems are increasingly heterogeneous, typically combining CPUs with accelerators such as GPUs. A programming model that enables workloads to be efficiently distributed across different hardware architectures is therefore important.


Task-based programming models address these challenges by embedding dynamic schedulers within the runtime system, allowing \textit{tasks}, i.e., units of work that can be executed in parallel, to be automatically mapped onto the underlying hardware resources. Such models have proven effective in addressing both asynchronous scheduling~\cite{buttari2009class, agullo2014task} and heterogeneous scheduling~\cite{agullo2012hybridization, agullo2017achieving} challenges in the optimization of numerical applications.

In addition to the asynchronicity inherent in certain numerical algorithms, hardware resource contention and operating system management introduce non-deterministic latencies in many-core systems~\cite{hoefler2010characterizing, hood2010performance}.  Consequently, even in a simple parallel program, threads may require different amounts of time to complete the same workload. Such variability can lead to excessive waiting in the fork-join model, whereas task-based programming models can better tolerate and mitigate these effects.

In this paper, we investigate the effectiveness of task-based programming models in coping with the asynchronicity introduced by many-core systems. As a benchmark application, we consider the solve of the 2D Poisson equation using the red-black Gauss–Seidel method. This benchmark is simple yet representative of a broad class of stencil applications commonly found in scientific computing. Although it is highly homogeneous from an algorithmic perspective, its execution on hardware is not expected to be perfectly regular, as variability is introduced by the machine.  We evaluate two task-based implementations using two programming models, namely OpenMP implemented in the GCC toolchain (hereafter referred to simply as OpenMP) and OmpSs-2~\cite{perez2017improving}, and compare them against a conventional fork-join implementation based on the OpenMP \texttt{parallel for} construct. The evaluation is conducted on two systems: JUWELS (x86) and HAICGU (Arm).

Previous studies on task-based stencil applications~\cite{boillot2014task, heller2013application} have generally concluded that task-based approaches provide limited benefits for regular workloads and are therefore less attractive than conventional fork-join parallelization strategies. In contrast, our results show that task-based models can remain competitive for regular stencil applications, particularly on modern many-core systems where architectural asynchronicity becomes increasingly important. In particular, the OmpSs-2 implementation, benefiting from its NUMA-aware support~\cite{maronas2023mitigating}, outperforms the OpenMP tasking implementation (hereafter referred to as OpenMP task) and achieves performance comparable to the OpenMP \texttt{parallel for} implementation. Moreover, on HAICGU, which features a large number of cores and NUMA (non-uniform memory access) domains, OmpSs-2 delivers both higher and more stable performance than OpenMP \texttt{parallel for}.

The remainder of this paper is organized as follows. Section~\ref{sec:background} introduces the benchmark implementations, Section~\ref{sec:spec} describes the experimental setup, Section~\ref{sec:eval} presents the performance results and analysis, and Section~\ref{sec:conclude} concludes the paper with a summary and directions for future work.



\section{Benchmark Design and Parallelization}
\label{sec:background}


The 2D Poisson equation is an elliptic partial differential equation (PDE) widely used in physics and engineering (e.g., electrostatics, fluid dynamics, and heat conduction). It is also a well-studied kernel in the context of high-performance computing (HPC), as it can exemplify behaviors of a wide range of stencil applications. The equation with Dirichlet boundary conditions can be expressed as
\[
\nabla^2 u = \frac{\partial^2 u}{\partial x^2} + \frac{\partial^2 u}{\partial y^2} = f(x, y) \text{ for } (x,y) \in \Omega, \enspace u(x,y) = g(x,y) \text{ for } (x,y) \in \partial \Omega,
\]
where $\Omega$ is a suitable domain.

To solve this equation numerically, we discretize $\Omega$ into a uniform grid with spacing $\Delta x$ and $\Delta y$. For simplicity, we assume a square grid where $\Delta x = \Delta y = h = \tfrac{1}{N+1}$. The continuous function $u(x, y)$ is approximated at discrete grid points as $u_{i,j} = u(i\Delta x, j\Delta y)$. Applying the central difference approximation to the second derivatives yields the standard 5-point stencil:
\begin{equation} \label{eq:update}
u_{i,j} = \frac{1}{4} \left( u_{i+1,j} + u_{i-1,j} + u_{i,j+1} + u_{i,j-1} - h^2 f_{i,j} \right), i,j = 1,\ldots,N, 
\end{equation}
where $u_{i\pm 1,j}, u_{i,j\pm 1}$ take the boundary values $g((i\pm 1)\Delta x,j\Delta y), g(i\Delta x, (j\pm 1)\Delta y)$ if $i\pm 1, j\pm 1 \in \{0,N+1\}$.

The resulting linear system can be solved iteratively using the Gauss–Seidel method, where each grid point is updated using the most recently available values of its neighboring points. To expose parallelism, we employ the red-black Gauss–Seidel variant, which partitions the grid into two sets of points, labeled red and black, following a checkerboard pattern. Since points of the same color do not directly depend on one another, all red or all black points can be updated concurrently without data conflicts. In this paper, grid points satisfying $i+j$ even are labeled red, while the remaining points are labeled black.

In the following, we first present a simple OpenMP \texttt{parallel for} implementation to introduce the iterative method. We then transition to an overview of task-based programming, exploring the historical evolution, core motivations, and various models that define this paradigm. Finally, we present our task-based implementations of the benchmark.

\subsection{OpenMP Parallel For Implementation}

We show our \texttt{parallel for} implementation of the red-black Gauss–Seidel method for solving the 2D Poisson equation in Algorithm~\ref{alg:rb_gs_parallel}. We consider a square grid of size $n \times n$ with Dirichlet boundary conditions. We store the values associated with red and black sites separately. Specifically, the value of $u_{i,j}$ is stored in \texttt{grid\_red} when $i+j$ is even, and in \texttt{grid\_black} otherwise. The same storage scheme is used for the source function $f$.

\begin{algorithm}[btp]
\caption{Parallel For Red-Black Gauss-Seidel}
\label{alg:rb_gs_parallel}

\SetKwInOut{Input}{Require}
\SetKwInOut{Output}{Ensure}

\Input{$\text{grid\_red}$, $\text{grid\_black}$, $\text{f\_red}$, $\text{f\_black}$, $h$, $n$, $\text{num\_iterations}$}
\Output{$\text{grid\_red}$, $\text{grid\_black}$ (Converged numerical solution)}
\BlankLine
\texttt{\textbf{\#pragma omp parallel}}

\For{$it = 0; \ it < \text{num\_iterations} - 1; \ it++$}{
    \tcc{\text{STEP 1: Update red sites}}
    \texttt{\textbf{\#pragma omp for}}
    
    \For{$x = 1; \ x < n - 1; \ x++$}{
        $y\_start \leftarrow (x \% 2 == 0) ? 2 : 1$\;
        \For{$y = y\_start; \ y < n - 1; \ y += 2$}{
            $\text{grid\_red}_{x,y} \leftarrow \frac{1}{4}(\text{grid\_black}_{x+1,y}  + \text{grid\_black}_{x-1,y} + \text{grid\_black}_{x,y+1} + \text{grid\_black}_{x,y-1} - h^2 \text{f\_red}_{x,y})$\;
        }
    }

    \BlankLine
    \tcc{\text{STEP 2: Update black sites}}
    \texttt{\textbf{\#pragma omp for}}
    
    \For{$x = 1; \ x < n - 1; \ x++$}{
        $y\_start \leftarrow (x \% 2 == 0) ? 1 : 2$\;
        \For{$y = y\_start; \ y < n - 1; \ y += 2$}{
            $\text{grid\_black}_{x,y} \leftarrow \frac{1}{4}(\text{grid\_red}_{x+1,y}  + \text{grid\_red}_{x-1,y} + \text{grid\_red}_{x,y+1} + \text{grid\_red}_{x,y-1} - h^2 \text{f\_black}_{x,y})$\;
        }
    }
}
\end{algorithm}

\SetKwProg{Fn}{Function}{ is}{end}
\begin{algorithm}[btp]
\caption{OpenMP Task Red-Black Gauss-Seidel}
\label{alg:rb_gs_task}
\SetKwInOut{Input}{Require}
\SetKwInOut{Output}{Ensure}

\Input{$\text{grid\_red}$, $\text{grid\_black}$, $bs$ (block size), $nb$ (number of blocks), $\text{num\_iterations}$}
\Output{$\text{grid\_red}$, $\text{grid\_black}$}
\BlankLine
\texttt{\textbf{\#pragma omp parallel}}\;
\texttt{\textbf{\#pragma omp single}}\;
\For{$it = 0; \ it < \text{num\_iterations}; \ it++$}{
    $\text{update\_color}(\text{grid\_red}, \text{grid\_black}, bs, nb, \dots)$ \tcp*[r]{update red}
    
    $\text{update\_color}(\text{grid\_black}, \text{grid\_red}, bs, nb, \dots)$ \tcp*[r]{update black}
}

\BlankLine

\Fn{\text{update\_color}($\text{grid\_self}$, $\text{grid\_neighbor}$, $bs$, $nb$ , $\dots$ )}{
    \BlankLine
    \tcc{\text{STEP 1: Update boundary block 0}}
    \texttt{\textbf{\#pragma omp task depend(inout: grid\_self[0]) depend(in: grid\_neighbor[0], grid\_neighbor[bs])}}\;
    
    $\text{update\_block}(\text{grid\_self}[0], \text{grid\_neighbor}[0], \text{NULL}, \text{grid\_neighbor}[bs], \dots)$\;

    \BlankLine
    \tcc{\text{STEP 2: Update inner blocks}}
    \For{$b = 1; \ b < nb - 1; \ b++$}{
        \texttt{\textbf{\#pragma omp task depend(inout: grid\_self[b*bs]) depend(in: grid\_neighbor[b*bs], grid\_neighbor[(b-1)*bs], grid\_neighbor[(b+1)*bs])}}\;
        
        $\text{update\_block}(\text{grid\_self}[b * bs], \text{grid\_neighbor}[b * bs], \text{grid\_neighbor}[(b-1) * bs], \text{grid\_neighbor}[(b+1) * bs], \dots)$\;
    }

    \BlankLine
    \tcc{\text{STEP 3: Update boundary block nb-1}}
    \texttt{\textbf{\#pragma omp task depend(inout: grid\_self[(nb-1)*bs]) depend(in: grid\_neighbor[(nb-1)*bs], grid\_neighbor[(nb-2)*bs])}}\;
    
    $\text{update\_block}(\text{grid\_self}[(nb-1) * bs], \text{grid\_neighbor}[(nb-1) * bs], \text{grid\_neighbor}[(nb-2) * bs], \text{NULL}, \dots)$\;
}
\tcc{\text{update\_block} parameters mapping:}
\tcc{1st: Target block (current color)}
\tcc{2nd: Target block (opposite color)}
\tcc{3rd: Top neighbor block (opposite color)}
\tcc{4th: Bottom neighbor block (opposite color)}
\end{algorithm}

Since this implementation follows the fork-join model, it relies on bulk synchronization through barriers. The two \texttt{for} constructs in Lines 3 and 8 each introduce an implicit barrier at the end of their execution regions. These barriers ensure that all red-site updates are completed before the computation proceeds to the black-site updates, and repeating across all iterations.

\subsection{Task-Based Programming}

An early work on task-based programming was introduced in Cilk \cite{blumofe1995cilk}, where a program was considered a collection of \textit{procedures}. The procedure, later termed task, is a sequence of instructions that can be executed in parallel and managed by a scheduler to optimize performance on multiprocessor systems. This approach to parallelism can handle irregular control structures such as while-loops and recursive functions, as opposed to standardized for-loop parallelism supported by OpenMP in the 1990s. 
Several prominent task-based programming models emerged in the 2000s, including Intel Threading Building Blocks (TBB)~\cite{willhalm2008putting} and OpenMP with its tasking extensions~\cite{ayguade2008design}. Later, the HPX C++ library was introduced~\cite{kaiser2014hpx}, further expanding the ecosystem of task-based programming.

Task-based programming models were subsequently extended beyond shared-memory systems. StarPU was among the pioneering task-based runtime systems that provided an efficient interface to execute parallel tasks across heterogeneous hardware architectures~\cite{augonnet2009starpu}.  OmpSs introduced a task-based programming model supporting both homogeneous and heterogeneous systems~\cite{bueno2011productive}, adopting a design philosophy akin to OpenMP’s approach to shared-memory parallelism. Subsequent enhancements extended OmpSs to incorporate GPU acceleration~\cite{bueno2012productive}. This model has recently been developed into OmpSs-2.

In this paper, we evaluate the impact of the task-based execution on the red-black Gauss–Seidel method using two task-based programming models: OpenMP and OmpSs-2. These models are chosen because they follow a directive-based programming approach, where parallelism is expressed through compiler directives (pragmas) embedded in the source code rather than through explicit thread management. 


Although this work focuses on native task-based programming models, similar asynchronous execution strategies can also be implemented using distributed-memory approaches such as MPI. In MPI-based implementations, workloads are distributed across ranks while synchronization and communication between neighboring ranks are managed explicitly by the programmer. Compared with task-based programming models considered here, this approach generally requires substantially more programming effort, including manual management of synchronization and memory buffers.

\subsection{OpenMP Task and OmpSs-2 Implementations}

\begin{figure}
    \centering
    \includegraphics[width=0.3\linewidth]{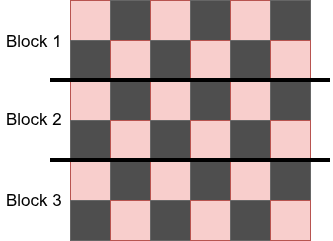}
    \caption{Illustration of the 2D grid data decomposition.}
    \label{fig:decomp}
\end{figure}

A task-based implementation of the 2D Poisson solver using red-black Gauss-Seidel requires defining tasks as units of work to be executed in parallel. To achieve better control over task granularity, a common approach is to decompose the data. For simplicity, we partition the 2D grid along one dimension, resulting in horizontal stripes, as illustrated in Fig.~\ref{fig:decomp}. 

In our task-based implementations, updating all red or black sites within a block constitutes a task, and we treat the number of blocks ($nb$) as an input. Thus, a total of $2 \cdot nb$ tasks are issued per iteration. Consider Fig.~\ref{fig:decomp} as an example, where the grid size is $6 \times 6$ and we set $nb = 3$; hence, $6$ tasks are issued in each iteration of the iterative method.

Algorithm~\ref{alg:rb_gs_task} illustrates our OpenMP task implementation. The core function, \texttt{update\_color()}, updates sites of a given color within a block. Data dependencies are specified to constrain task execution. In particular, an update requires read and write access to the block of its own color, as well as read access to the same block and the neighboring blocks of the opposite color. These dependencies are enforced through the \texttt{depend} clauses in Lines 7, 10, and 12. The implementation details of the \texttt{update\_block()} function are omitted from the algorithm for brevity, as it simply applies~\eqref{eq:update} to all red or black sites within the block.

Task-based programming models with data dependency support, such as OpenMP, enforces execution order through data tracking. More specifically, tasks that exhibit conflicting dependencies, such as read-after-write (RAW), write-after-read (WAR), or write-after-write (WAW) on the same data region, are serialized by the runtime system to preserve correctness. Hence, the correctness of Algorithm~\ref{alg:rb_gs_task} is ensured by issuing all red tasks first, followed by all black tasks (Lines 1-5). It is also worth noting that a \texttt{single} construct is used so that task generation is performed by only one thread.

The OmpSs‑2 implementation is similar to that of OpenMP. Changes are required for each pragma, i.e., from \texttt{\#pragma omp} to \texttt{\#pragma oss}. OmpSs‑2 initiates a thread pool at the beginning of the program; therefore it does not need a \texttt{parallel} construct. By default, one thread executes the main function for task creation, and all other runtime threads can be used for task execution. A \texttt{taskwait} construct is needed at the end of the iterations (i.e., after Line 5) to enforce a barrier so that all tasks are finished before exiting the region of interest (ROI).


\section{Experimental Setup}
\label{sec:spec}

Before presenting the evaluation of the performance of our implementations on hardware, we outline the test setup on JUWELS (x86) and HAICGU (Arm). The details of the hardware and software used on the two platforms are listed in Table~\ref{tab:spec}. We evaluate two task-based programming models: OmpSs-2 with the Nanos6~\cite{nanos6_bsc} v4.3 runtime and GCC OpenMP (GCC versions listed in Table~\ref{tab:spec}).

\begin{table}[btp]
    \centering
    \caption{Node-level hardware and software for benchmarking platforms.}
    \begin{tabular}{l  l  l}  
\toprule
& JUWELS & HAICGU \\
\midrule
Architecture & x86 & Arm\\
Processor  &   2$\times$Xeon Platinum 8168   &  2$\times$Kunpeng 920    \\
Number of Cores &  $2\times 24$ & $2\times 64$\\
Number of NUMA domains & 2 & 4\\
CPU Frequency   &   \qty{2.7}{\giga\hertz}   &  \qty{2.6}{\giga\hertz}  \\
L1 Data Cache & \qty{32}{\kibi\byte} per core  & \qty{64}{\kibi\byte} per core\\
L2 Cache & \qty{1}{\mebi\byte} per core & \qty{512}{\kibi\byte} per core\\
L3 Cache & \qty{33}{\mebi\byte} per CPU & \qty{32}{\mebi\byte} per CPU\\
SIMD Width & \qty{512}{\bit} & \qty{128}{\bit} \\
Memory & \qty{96}{\gibi\byte} & \qty{128}{\gibi\byte} \\
Theor.~Mem.~Bandwidth   &  \qty{256}{\giga\byte\per\second}   &  \qty{341}{\giga\byte\per\second}   \\
Meas.~Mem.~Bandwidth & \qty{155}{\giga\byte\per\second}   & \qty{218}{\giga\byte\per\second} \\
Roofline Ridge Point & \qty{16.2}{\flop\per\byte} & \qty{1.96}{\flop\per\byte}\\
Compiler Used & GCC v13.3.0 & GCC v14.1.0 \\
\midrule[\heavyrulewidth]
\bottomrule
\end{tabular}
\label{tab:spec}
\end{table}

In addition to the theoretical memory bandwidth, we use the STREAM benchmark~\cite{mccalpin1995memory} v5.10 to measure the peak attainable memory bandwidth. The values reported in Table~\ref{tab:spec} are obtained using the maximum available cores. Specifically, we use the STREAM Triad rates, which typically reflect the highest sustainable bandwidth on modern architectures. 

To assess the attainable performance of the benchmark, we employ the roofline model~\cite{williams2009roofline}. The roofline model characterizes application performance in terms of its arithmetic intensity, $\Ifp/\Imem$, where $\Ifp$ and $\Imem$ denote the floating-point operation count and memory traffic, respectively. For the benchmark considered in this work, the arithmetic intensity is approximately $0.29$.

Applications with low arithmetic intensity are memory-bound, meaning their performance is limited by memory bandwidth, whereas applications with high arithmetic intensity are compute-bound and limited by floating-point throughput. The transition between these two regimes occurs at the ridge point of the roofline model, an arithmetic-intensity threshold defined as the ratio of peak floating-point performance to peak memory bandwidth. Since the arithmetic intensity of the benchmark lies well below the ridge points of both platforms listed in Table~\ref{tab:spec}, the benchmark is firmly memory-bound. Consequently, its attainable performance can be estimated from the measured memory bandwidth as
\begin{equation} 
\label{eq:roofline} 
\text{Att. Perf.} = \text{Meas. Mem. Bandwidth} \times \frac{\Ifp}{\Imem}.
\end{equation}

We consider single-node execution, as OpenMP does not natively support distributed-memory parallelism. For distributed computing, MPI remains the de facto standard, and both programming models support a hybrid MPI+tasking approach. We also focus on CPU architectures. 

All performance results reported in this paper are obtained by executing the ROI ten times and recording the median value to reduce the impact of statistical noise and system interference. For simplicity, we use a 1:1 mapping between physical cores and runtime threads, and all available cores are used for all experiments. For OpenMP, thread affinity is configured using \texttt{OMP\_PROC\_BIND=spread} and \texttt{OMP\_PLACES=cores}. For OmpSs-2, thread pinning is performed using \texttt{taskset}, and NUMA tracking is enabled for the Nanos6 runtime.
\section{Evaluation}
\label{sec:eval}

\begin{figure}[btp]%
    \centering
    \includegraphics[width=0.8\textwidth]{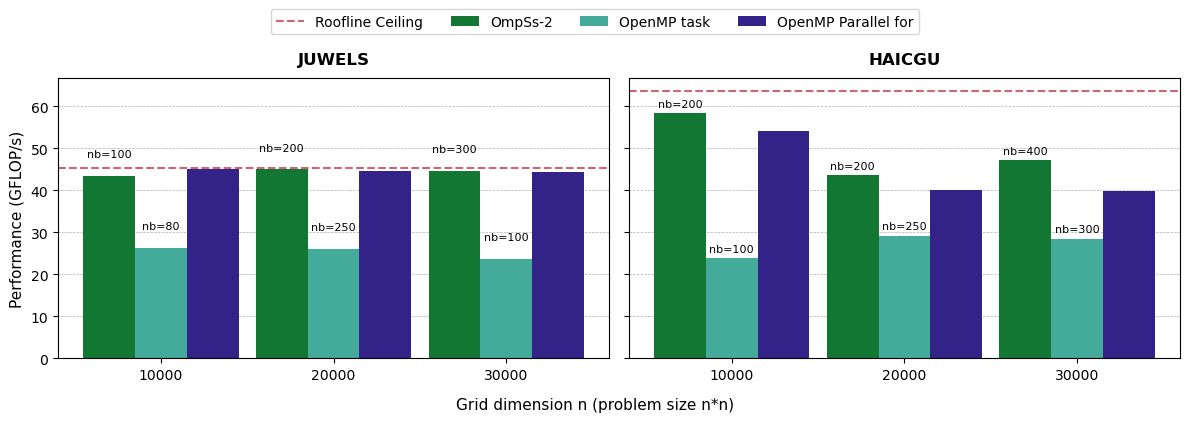}
    \caption{Performance (median) of red-black Gauss-Seidel with varying matrix size, 100 iterations are taken.}%
    \label{fig:perf}%
\end{figure}

We compare the implementations introduced in Section~\ref{sec:background}, namely the OpenMP \texttt{parallel for} baseline and the two task-based implementations: OpenMP task and OmpSs-2. Their performance results, together with the peak attainable performance derived from \eqref{eq:roofline}, are presented in Fig.~\ref{fig:perf}. We consider three problem sizes, each sufficiently large to ensure memory-bound execution, and fix the number of red-black Gauss-Seidel iterations to 100.

Compared with the OpenMP \texttt{parallel for} implementation, the task-based implementations introduce an additional tuning parameter, namely $nb$ in Algorithm~\ref{alg:rb_gs_task}, which controls task granularity. A small number of tasks may lead to hardware underutilization, while an excessively large number of tasks can incur significant runtime overhead. Through empirical analysis, we determine the task counts that achieve optimal performance, and report them in Fig.~\ref{fig:perf}.

The superior performance of OmpSs-2 on both architectures is attributable to its NUMA-aware support~\cite{maronas2023mitigating}.
NUMA is a widely used memory architecture in HPC systems that enables larger core counts and memory capacities by partitioning memory into multiple NUMA domains. However, this design introduces differences in memory access latency and bandwidth between local and remote NUMA domains.
Modern operating systems are aware of the NUMA topology and typically follow a first-touch policy, allocating memory in the NUMA domain of the thread that first accesses it. Consequently, if threads executing tasks are not carefully scheduled on the NUMA domains where the corresponding data reside, remote memory accesses may occur, resulting in higher memory transfer costs and degraded performance.

Working with operating systems' first-touch policy is relatively straightforward with OpenMP \texttt{parallel for}: the data can simply be initialized using another \texttt{parallel for} with the same access pattern as the subsequent computation. OmpSs-2 provides low-level memory allocation APIs that allow data to be distributed across NUMA domains in an interleaved manner, effectively mitigating the first-touch policy. Its task scheduler can also place tasks on the NUMA domains where their data reside through data tracking mechanisms. OpenMP task, however, does not provide such NUMA-aware support. Without enabling the NUMA-aware features of OmpSs-2, we observe performance degradations of approximately 40\% on JUWELS and 43\% on HAICGU. 

The NUMA effect is expected to be more pronounced on HAICGU than on JUWELS. The Kunpeng 920 processor used in HAICGU consists of four NUMA domains, within each of which the cores are organized in a ring topology. This design likely incurs higher synchronization and communication costs between cores compared with the Intel Xeon Platinum 8168 used in JUWELS, which connects its cores via a two-dimensional mesh on-chip network and has two NUMA domains. When data placement does not match the data access pattern (here achieved by using numactl~\cite{numactl_man} to interleave memory allocation), we observe that the STREAM benchmark reports bandwidth reductions of 47\% and 55\% relative to the values listed in Table~\ref{tab:spec} for JUWELS and HAICGU, respectively.

\begin{figure}[btp]%
    \centering
    \includegraphics[width=0.8\textwidth]{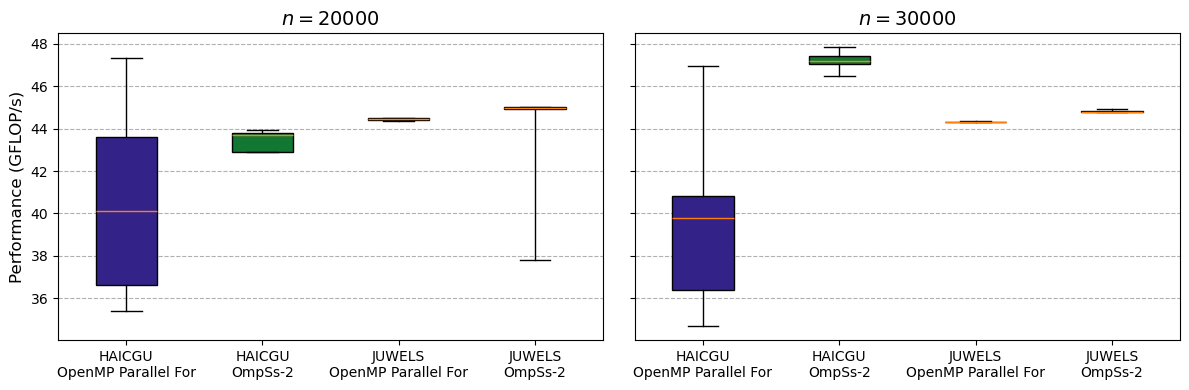}
    \caption{Box-and-whisker plots (five-number summary: min, first quartile, median, third quartile, and max) of performance.}%
    \label{fig:perf_var}%
\end{figure}

OmpSs-2 achieves performance comparable to OpenMP \texttt{parallel for} on JUWELS, and outperforms it on HAICGU. Furthermore, for large problem sizes on HAICGU, \texttt{parallel for} exhibits greater execution variability than OmpSs-2, as shown in Fig.~\ref{fig:perf_var}. Although some runs of OpenMP \texttt{parallel for} can achieve performance similar to OmpSs-2 on HAICGU, its overall performance is less stable. 

Such instability may arise from varying execution latencies that force threads to wait at barriers. These latencies can be caused by operating system noise or resource contention, where threads compete for shared resources. They are more likely to occur on the Kunpeng 920 processor of HAICGU as it features substantially more cores and NUMA domains. To further investigate this behavior, we use Extrae~\cite{extrae_bsc} to trace the OpenMP \texttt{parallel for} execution and Ovni~\cite{OVNI_BSC} to trace the OmpSs-2 execution. We further use Paraver~\cite{Paraver} to extract the execution traces.

\begin{figure}[btp]%
    \centering
    \subfloat[\centering First 10 iterations]{{\includegraphics[width=0.4\textwidth]{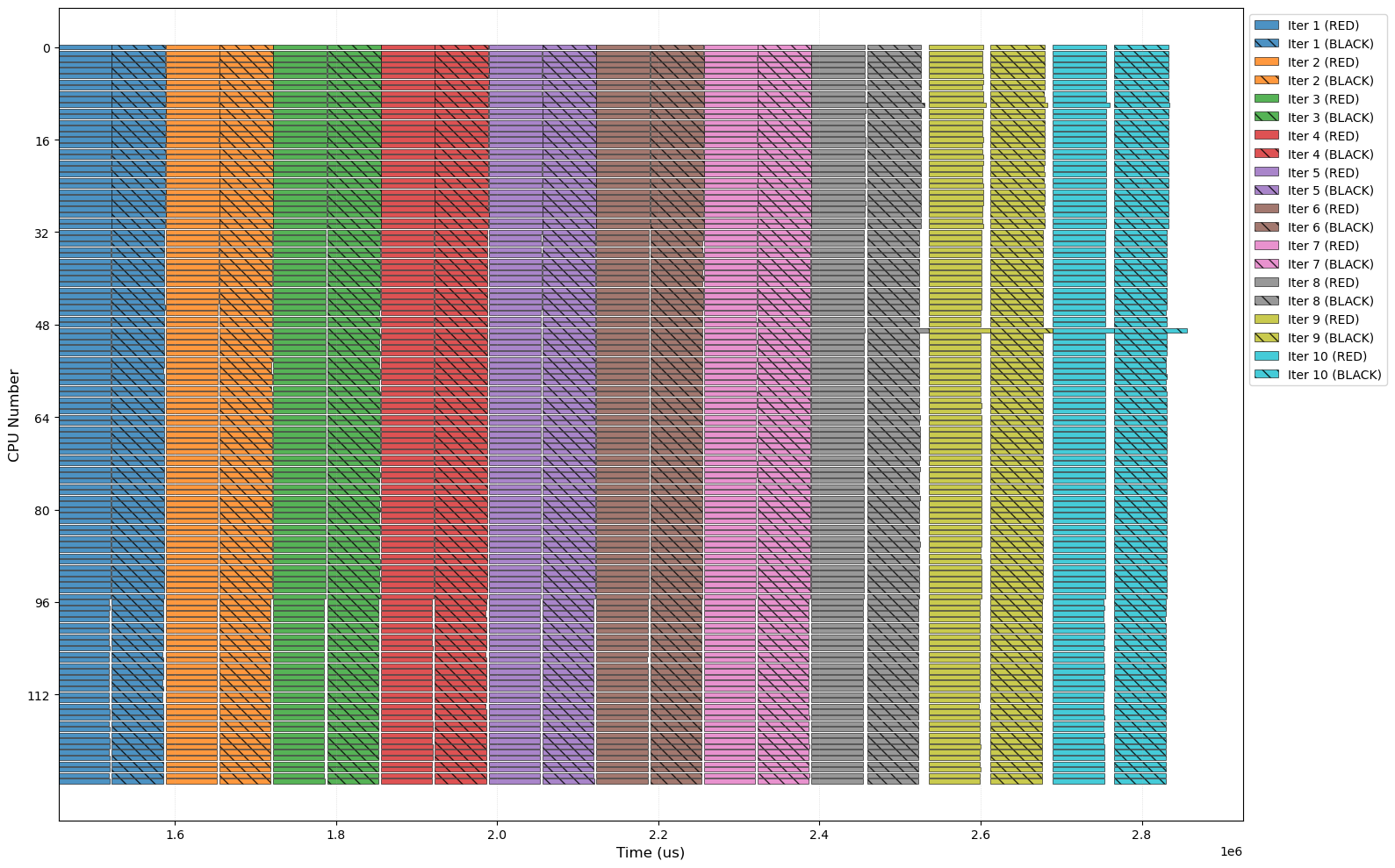} }}%
    \subfloat[\centering Entire 100 iterations]{{\includegraphics[width=0.4\textwidth]{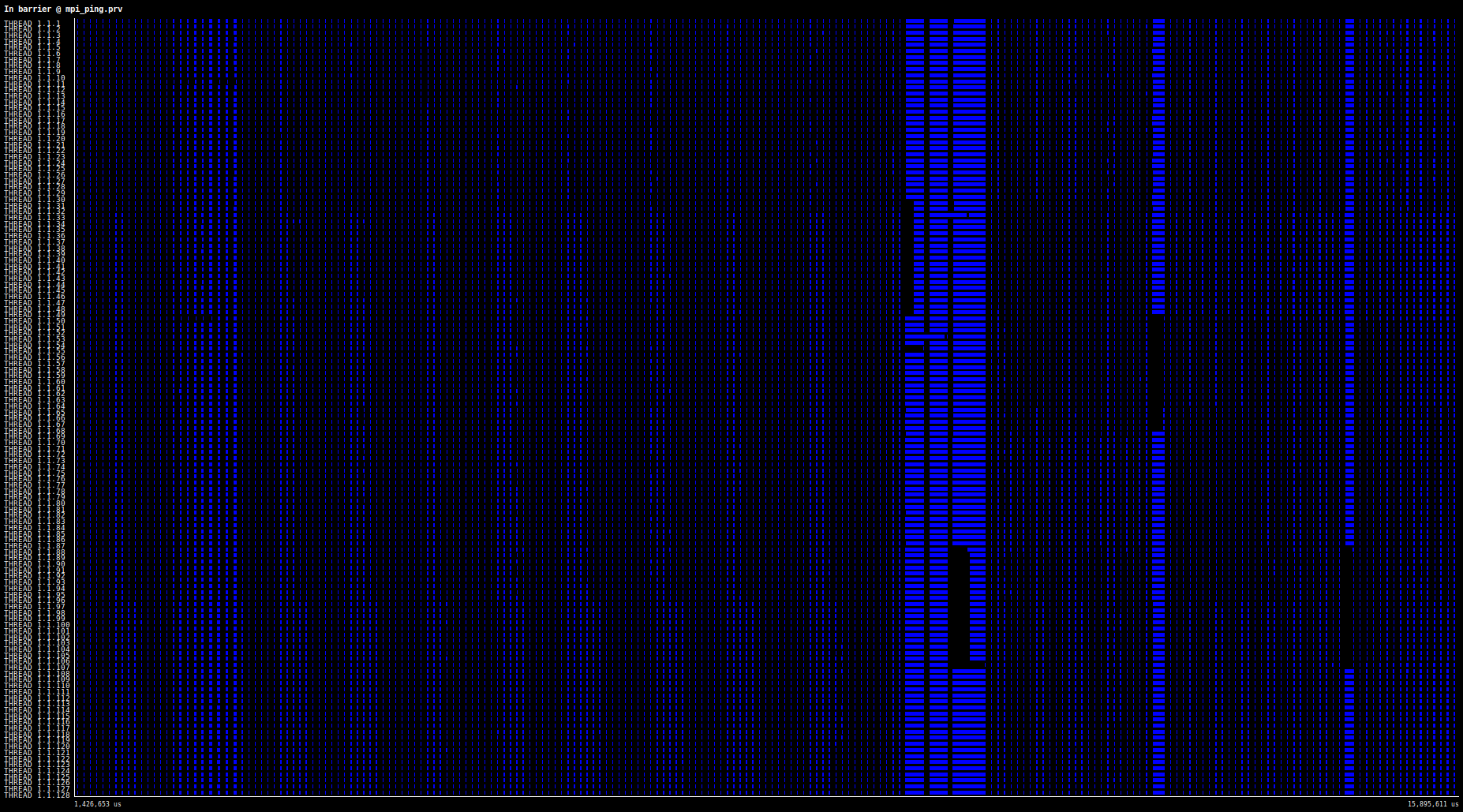} }}%
    \caption{Mapping of OpenMP execution to CPU cores on HAICGU ($n=30\,000$). 
    }%
    \label{fig:paraver}%
\end{figure}

We plot the first 10 iterations of an OpenMP \texttt{parallel for} execution in Fig.~\ref{fig:paraver}(a). A clear separation between phases corresponding to different color updates can be observed, as all threads begin executing the same amount of work simultaneously, and the next phase starts only after all threads have completed their work. Execution variability is also evident, with some threads taking noticeably longer to finish than others, leaving the remaining threads idle and making the gaps between phases more pronounced. These gaps persist throughout the execution and are shown in blue in Fig.~\ref{fig:paraver}(b). 

\begin{figure}[btp]%
    \centering
    \subfloat[\centering JUWELS ($nb=300$)]{{\includegraphics[width=0.4\textwidth]{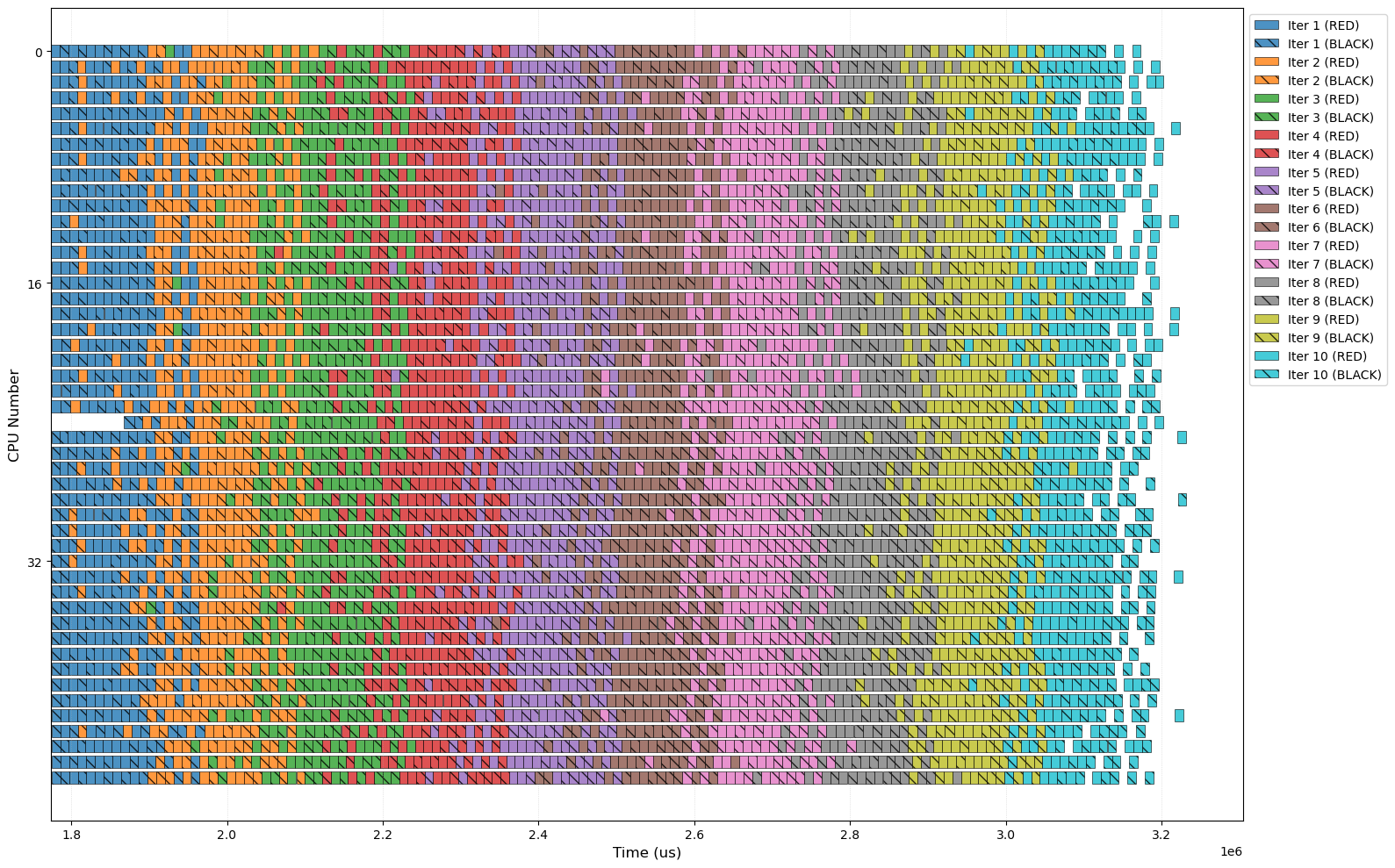} }}%
    \subfloat[\centering HAICGU ($nb=400$)]{{\includegraphics[width=0.4\textwidth]{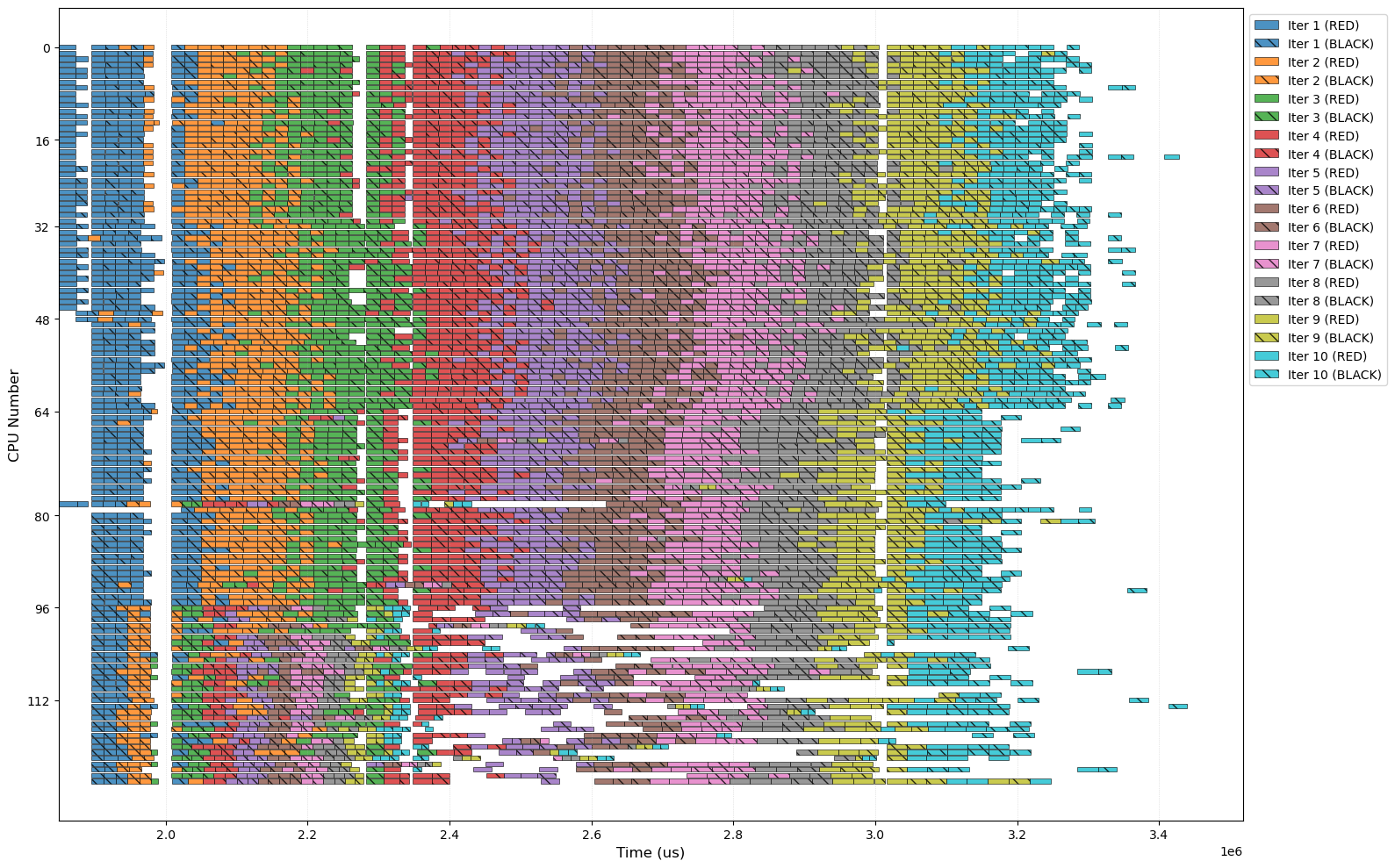} }}%
    \caption{Mapping of OmpSs-2 execution of the first 10 iterations to CPU cores on HAICGU ($n = 30\,000$).}%
    \label{fig:breakdown}%
\end{figure}

Such clear separation between phases is not required in task-based implementations, as they do not rely on rigid synchronization between all red and all black updates. Overlaps between red and black phases, as well as between successive iterations, can be observed in the OmpSs-2 execution traces shown in Fig.~\ref{fig:breakdown}. The threads remain continuously busy throughout execution. Note that the gaps in Fig.~\ref{fig:breakdown}(b) for cores 96--128 do not indicate idle cores; rather, they correspond to tasks from later iterations being executed. This ability to hide execution variability through dynamic scheduling explains the higher and more stable performance of OmpSs-2 over OpenMP \texttt{parallel for} on HAICGU.

Another potential source of performance differences is cache reuse. While a \texttt{parallel for} implementation follows a regular work distribution, tasks in a task-based implementation are scheduled dynamically by the runtime system. Consequently, the resulting data access pattern depends on task granularity, scheduling decisions, and hardware characteristics, leading to different cache locality properties. 

However, with the problem size and task granularity used in this work, each task accesses a sufficiently large amount of data that the relevant cache lines are unlikely to remain in cache until a core finishes its current task and is assigned another task by the runtime. We therefore do not expect the cache reuse of the task-based implementation to differ significantly from that of OpenMP \texttt{parallel for}.
This is confirmed by hardware performance counter measurements on JUWELS\footnote{L3 cache data-read misses and total data-read requests are obtained using the \texttt{OFFCORE\_RESPONSE} counter with \texttt{request=ALL\_DATA\_RD} and the filters \texttt{L3\_MISS} and \texttt{ANY\_RESPONSE}. L2 cache data-read misses and total data-read requests are obtained using the \texttt{L2\_RQSTS} counter, which includes information on both on-demand and hardware prefetcher data reads.}, which show nearly identical cache refill rates for OmpSs-2 and OpenMP \texttt{parallel for}. This is also consistent with the small performance difference observed between the two implementations on JUWELS.

\section{Conclusion}
\label{sec:conclude}

Although task-based paradigms have been widely introduced in numerical computing, previous studies have primarily focused on their ability to parallelize irregular workloads for performance optimization. In this work, we analyzed task-based implementations of the 2D Poisson solver using the red-black Gauss-Seidel method, and compared them against a conventional OpenMP \texttt{parallel for} implementation. Our results show that the additional implementation complexity of the task-based approach is modest and remains comparable to that of \texttt{parallel for}. Transitioning from a \texttt{parallel for} implementation to either OpenMP task or OmpSs-2 requires relatively little programming effort, particularly when compared with achieving a similar fine-grained parallelism using MPI.

Through experimental evaluation, we found that the best-performing task-based variant, OmpSs-2, achieves performance comparable to or better than \texttt{parallel for}. On HAICGU, OmpSs-2 delivers both higher and more stable performance. Its dynamic scheduling mechanism mitigates the impact of execution-time variability by keeping hardware resources occupied and reducing the synchronization overhead associated with the fork-join model. These results demonstrate that task-based approaches can effectively tolerate architectural asynchronicity, making them a competitive option even for regular workloads.

Future work includes extending this study from the simple 2D Poisson problem to more complex systems involving higher dimensions and more sophisticated computations. In particular, multi-color variants of Gauss-Seidel introduce more synchronization requirements, potentially increasing the benefits of task-based execution. It would also be interesting to investigate other iterative methods with behavior similar to red-black Gauss-Seidel, such as the odd-even techniques commonly used in scientific computing for various fields of study.



\begin{acknowledgement}
The authors gratefully acknowledge the Gauss Centre for Supercomputing e. V. for funding this project by providing computing time through the John von Neumann Institute for Computing (NIC) on the GCS Supercomputer JUWELS at J\"{u}lich Supercomputing Centre (JSC), under the project with id MUL-TRA.
Furthermore, we want to thank the Open Edge and HPC Initiative for access to an Arm-based development environment through the HAICGU cluster at the Goethe University of Frankfurt. 

Funding for parts of this work has been received from the European Union’s HORIZON MSCA Doctoral Networks programme AQTIVATE, under grant agreement No. 101072344. 
G.R-H. acknowledges financial support from the EoCoE-III project, which has received funding from the European High Performance Computing Joint Undertaking under grant agreement No. 101144014.
\ \\ \ \\

\end{acknowledgement}

\vspace{\baselineskip}
\bibliographystyle{plain}
\bibliography{ref} 

\end{document}